\documentclass[a4paper,11pt]{article}
\usepackage{pos}
\usepackage[normalem]{ulem} 

\title{Production and decay of a heavy boson in association with top
quarks}

\author*[a]{Ma\l{}gorzata Worek}
\notes{\note{TTK-22-27, P3H-22-076}}

\affiliation[a]{
Institute for Theoretical Particle Physics and Cosmology\\
  RWTH
  Aachen University, D-52056 Aachen, Germany}

\emailAdd{worek@physik.rwth-aachen.de}

\abstract{We  briefly report on NLO QCD corrections
to the $pp \to t\bar{t} B +X$ process, where $B$ stands for
$H,\,W^\pm$ or $Z$. In the calculation all double-, single
and non-resonant Feynman diagrams, interferences and finite-width
effects are included for the top quark and $W^\pm/Z$ gauge
bosons. Numerical results are shown at the differential cross 
section
level for various factorisation and renormalisation scale settings and
a few PDF sets. The main theoretical uncertainties 
associated with neglected higher-order terms in the perturbative
expansion and with different parametrisations of the PDFs are also
discussed. Furthermore, we compare our complete predictions for $pp
\to t\bar{t} B+X$ in multi-lepton final states to the calculation in
the narrow-width approximation.}

\FullConference{%
  Loops and Legs in Quantum Field Theory - LL2022,\\
  25-30 April, 2022\\
  Ettal, Germany
}

\begin{document}
\maketitle

\section{Introduction}

The associated production of a pair of top quarks with massive gauge
bosons in multi-lepton final states has been extensively scrutinised in
recent years. From the experimental point of view, an accurate study of
$pp \to t\bar{t}W$ and $pp\to t\bar{t}Z$  has become
feasible thanks to the increasing amount of data collected at the LHC
with $\sqrt{s}=13$ TeV by ATLAS and CMS 
\cite{CMS:2017ugv,ATLAS:2019fwo,CMS:2019too,CMS:2021aly,
ATLAS:2021fzm}. From the theoretical perspective, precise predictions
for $pp \to t\bar{t}W$ and $pp\to t\bar{t}Z$ are of particular interest
because both processes can receive sizeable contributions from
physics beyond the Standard Model (SM).  In addition, $pp \to
t\bar{t}Z$ is the most sensitive process to directly measure the
coupling of the top quark to the $Z$ gauge boson.  It gives an important
insight into top quark properties  that is complementary to
$t\bar{t}$ and single top quark production as well as to the $t\to Wb$
decay. Any deviations of the coupling strength of the top quark to the
$Z$ boson from its SM value might imply the existence of new physics
effects.  Furthermore, both $pp \to t\bar{t}Z$ and $pp \to t\bar{t}W$ are dominant
backgrounds to several searches for new physics as they comprise final
states with multiple charged leptons, missing transverse momentum
$(p_T^{miss})$ and
$b$-jets,  which are vigorously searched for at the LHC.  Last
but not
least, both processes play a prominent role in the measurements of the
SM Higgs boson in $t\bar{t}H$ production with $H\to ZZ\to 4\ell$,
$H\to W^+W^-\to 2\ell 2\nu$ and $H\to
\tau^+\tau^-$ decays. The $t\bar{t}H$ process, on the other hand, gives a direct window
to probe the top quark Yukawa coupling. Even though $pp \to t\bar{t}H$
production contributes only about $1\%$ to the total Higgs boson
production cross section, it was observed in 2018 by  ATLAS and CMS
\cite{CMS:2018uxb,ATLAS:2018mme}. In  recent
measurements of $t\bar{t}H$ and $t\bar{t}W^\pm$ production in
multi-lepton final states \cite{ATLAS:2019nvo,CMSnote}, the resulting
$t\bar{t}W^\pm$ normalisation has been found to be higher than the NLO
(QCD + EW) + NNLL theoretical prediction as provided in Ref.
\cite{Broggio:2016zgg,Kulesza:2018tqz,Broggio:2019ewu,Kulesza:2020nfh}.
Apart from the $t\bar{t}W^\pm$ normalisation, a tension in the
modelling of the final-state kinematics in the phase-space regions
dominated by the $t\bar{t} W^\pm$ process has been observed. This
tension could not be explained by multipurpose Monte Carlo
(MC) generators, which are currently employed by the ATLAS and CMS
collaborations. Consequently, the need for more precise theoretical
results for $t\bar{t} B$ production is now higher than ever.  Such theoretical
predictions should include higher order QCD corrections both to the
production and decays of top quarks and $W$ gauge bosons as well as
$t\bar{t}$ spin correlations at the same level of accuracy.

First next-to-leading order (NLO) QCD calculations for the $pp \to
t\bar{t}B+X$ process with $B=H,Z,W^\pm$  that meet the above mentioned 
conditions have been carried out in the narrow-width approximation
(NWA)
\cite{Campbell:2012dh,Rontsch:2014cca,Zhang:2014gcy,Bevilacqua:2020pzy,
Hermann:2021xvs,Stremmer:2021bnk,Bevilacqua:2022nrm}.  First full NLO
QCD computations, which include complete top quark and $W^\pm/Z$
off-shell effects for the $pp \to t\bar{t}B+X$ process in the
multi-lepton channel, have been provided in
Refs. \cite{Denner:2015yca,Bevilacqua:2019cvp,Denner:2020hgg,
Bevilacqua:2020pzy,Stremmer:2021bnk,Bevilacqua:2022nrm}.  In these
computations, off-shell top quarks as well as $W^\pm/Z$ gauge bosons
have been described by Breit-Wigner propagators. Furthermore, double-,
single- as well as non-resonant contributions along with all
interference effects have been consistently incorporated at the matrix
element level. In a few cases, even NLO electroweak (NLO EW)
corrections to the full off-shell $t\bar{t}B+X$ production have been
included \cite{Denner:2016wet,Denner:2021hqi}. A further step towards
a more precise modelling of the on-shell $t\bar{t}B+X$ production
process has been achieved by including the subleading electroweak
corrections \cite{Dror:2015nkp, Frederix:2017wme,Frederix:2020jzp} and
improving merging  procedures in the presence of one or two additional
jets. Finally, a great effort has been  made  to match $pp \to
t\bar{t}B+X$ production to various parton shower programs
Refs. \cite{Garzelli:2011vp,Garzelli:2011is,Garzelli:2012bn,
Hartanto:2015uka,Maltoni:2015ena,FebresCordero:2021kcc,
Bevilacqua:2021tzp,Ghezzi:2021rpc}.

In this contribution, we briefly summarise NLO QCD predictions for the
$pp \to t\bar{t}B+X$ process in multi-lepton decay channels. We
discuss theoretical results for this process based on full off-shell
calculations and the NWA approach. Both types of predictions have been
provided with the help of the \textsc{Helac-Nlo} MC package
\cite{Bevilacqua:2011xh} that is built around the
\textsc{Helac-Phegas} software \cite{Cafarella:2007pc}. The
\textsc{Helac-Nlo} program comprises \textsc{Helac-Dipoles}
\cite{Czakon:2009ss,Bevilacqua:2013iha}, which is used for the
calculation of the real emission contributions, as well as
\textsc{Helac-1Loop} \cite{vanHameren:2009dr}, which is employed for
the evaluation of the virtual corrections.

\section{Differential Cross-Section Results}

%
\begin{figure}[t!]
  \begin{center}
    \includegraphics[width=0.45\textwidth]{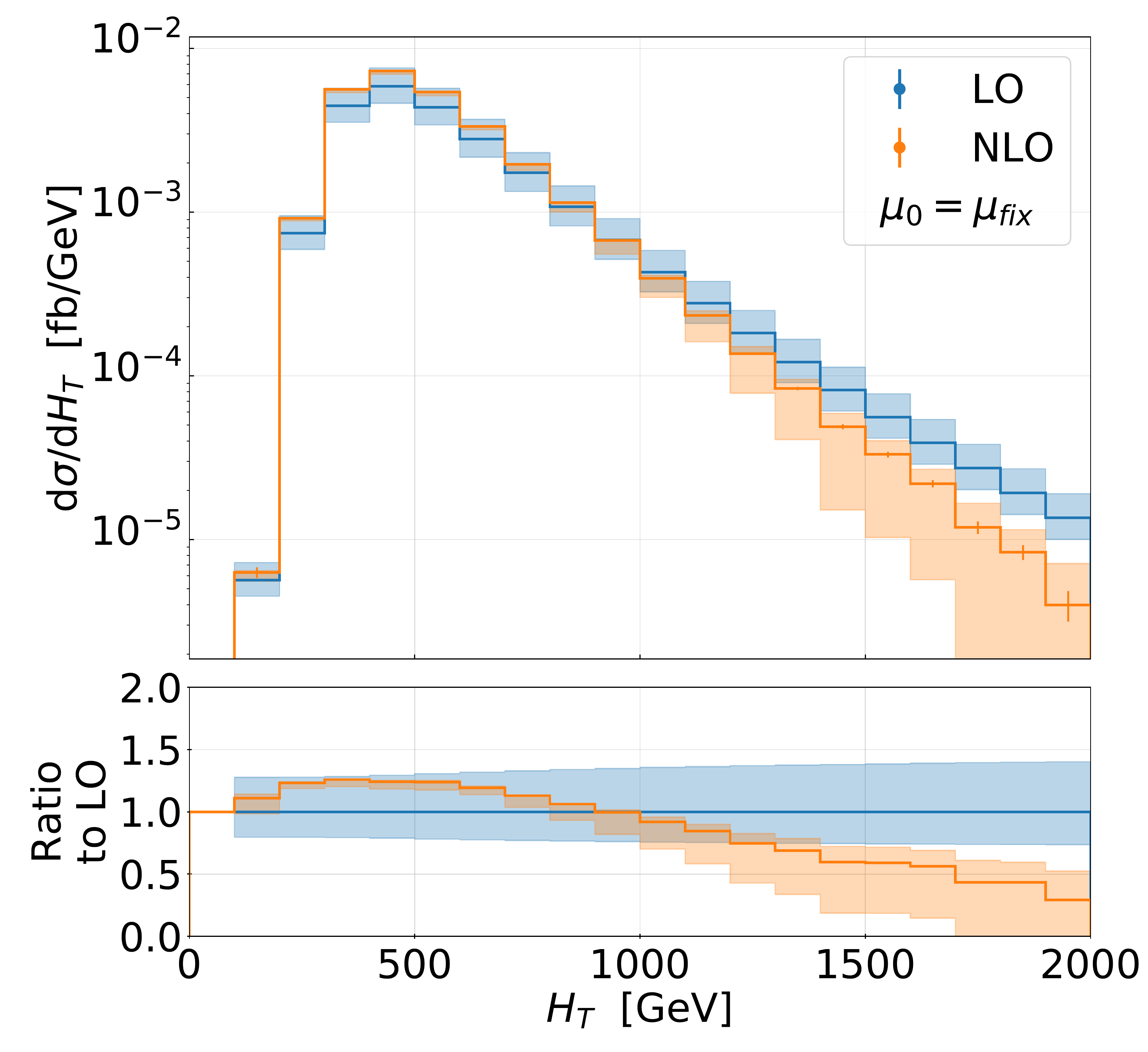}
   \includegraphics[width=0.45\textwidth]{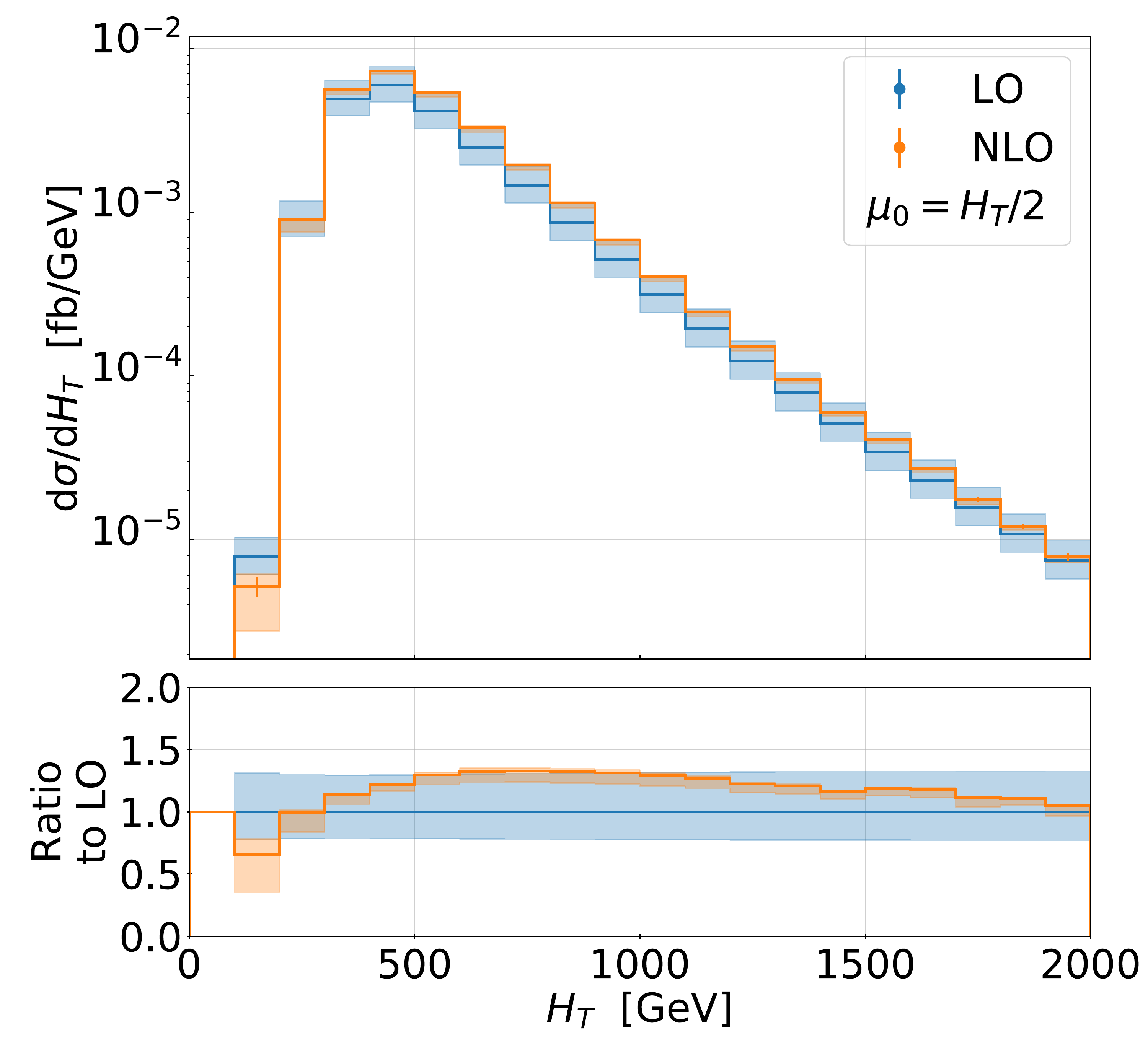}
\end{center}
\caption{\it \label{fig1}
Differential distributions at LO and NLO in QCD with the corresponding
uncertainty bands for the observable $H_T$ for the $pp \to e^+ \nu_e
\mu^- \bar{\nu}_\mu b\bar{b} H +X$ process at the LHC with
$\sqrt{s} = 13$ TeV. Results are presented for $\mu_0 = H_T /2$ and
$\mu_0 = \mu_{fix}$.  The corresponding scale uncertainty bands as well as Monte Carlo integration errors are also given. The lower
panels display the differential ${\cal K}$-factor.}
\end{figure}

In the following, we discuss various aspects of higher-order
predictions for the $pp \to t\bar{t}B+X$ process at the LHC  Run II
with $\sqrt{s}=13$ TeV. We start with assessing the size of the NLO
QCD corrections and the perturbative stability of the renormalisation
$(\mu_R)$ and factorisation $(\mu_F)$ scale settings.  In Figure
\ref{fig1}, we present differential distributions at LO and NLO QCD for
the $H_T$ observable defined as
$H_T=p_{T,\,b_1}+p_{T,\,b_2}+p_{T,\,e^+}+p_{T,\,\mu^-}+p_{T,\,miss}+p_{T,\,H}$
for the $pp \to e^+ \nu_e \mu^- \bar{\nu}_\mu b\bar{b} H +X$
process. Results are presented for $\mu_R=\mu_F=\mu_0$ where $\mu_0 =
\mu_{fix}= m_t+m_H /2$ (left panel) and $\mu_0=H_T /2$ (right
panel). Also shown are the corresponding uncertainty bands, which are
obtained using a $7$-point scale variation, as well as Monte Carlo
integration errors. The lower panels display the differential ${\cal
K}$-factor. The latter is defined according to ${\cal K}=\left( d
\sigma^{NLO}/dX\right)/\left(d\sigma^{LO}/dX\right)$, where
$X=H_T$. All input parameters, PDF sets and restrictions on the
kinematics on the final states are given in Ref. \cite{Stremmer:2021bnk}.

For the fixed scale choice that we employed, $\mu_0 = \mu_{fix}=
m_t+m_H /2$, we can observe that for some choices of the $\xi$
parameter, where $\mu_R = \mu_F = \xi \mu_0$, the NLO results become
negative. This happens in the high-energy tails of $H_T$. Furthermore,
in the same kinematic regions, scale variation bands at LO and NLO do
not overlap anymore. In addition, the scale variation at NLO actually
exceeds the scale variation of the LO predictions. At the differential
level a fixed scale choice cannot handle the dynamics of the process
in certain phase-space regions, leading to perturbative
instabilities. All these effects, however, might be accommodated by a
judicious choice of a dynamical scale setting. For example,
for $\mu_0=H_T/2$, scale uncertainties are well below 
$10\%$. In addition, we find that NLO predictions are completely
included in the uncertainty bands of the LO prediction. The higher
order QCD corrections are not constant in both cases and reach up to
$35\%$ even for the dynamical scale setting. Thus, the NLO QCD
corrections are necessary for a precise prediction for $H_T$ .
%
\begin{figure}[t!]
  \begin{center}
      \includegraphics[width=0.45\textwidth]{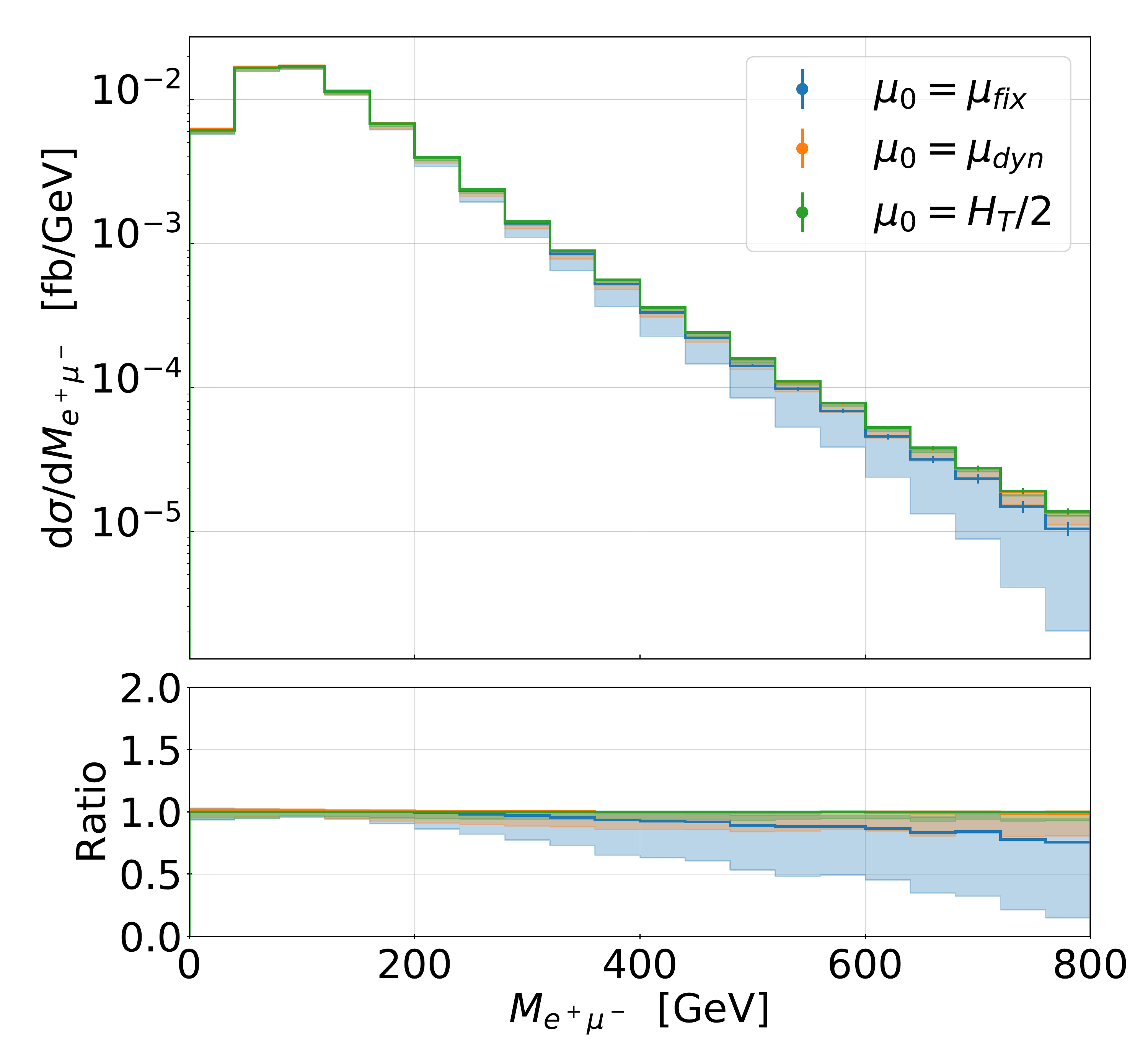}
  \includegraphics[width=0.45\textwidth]{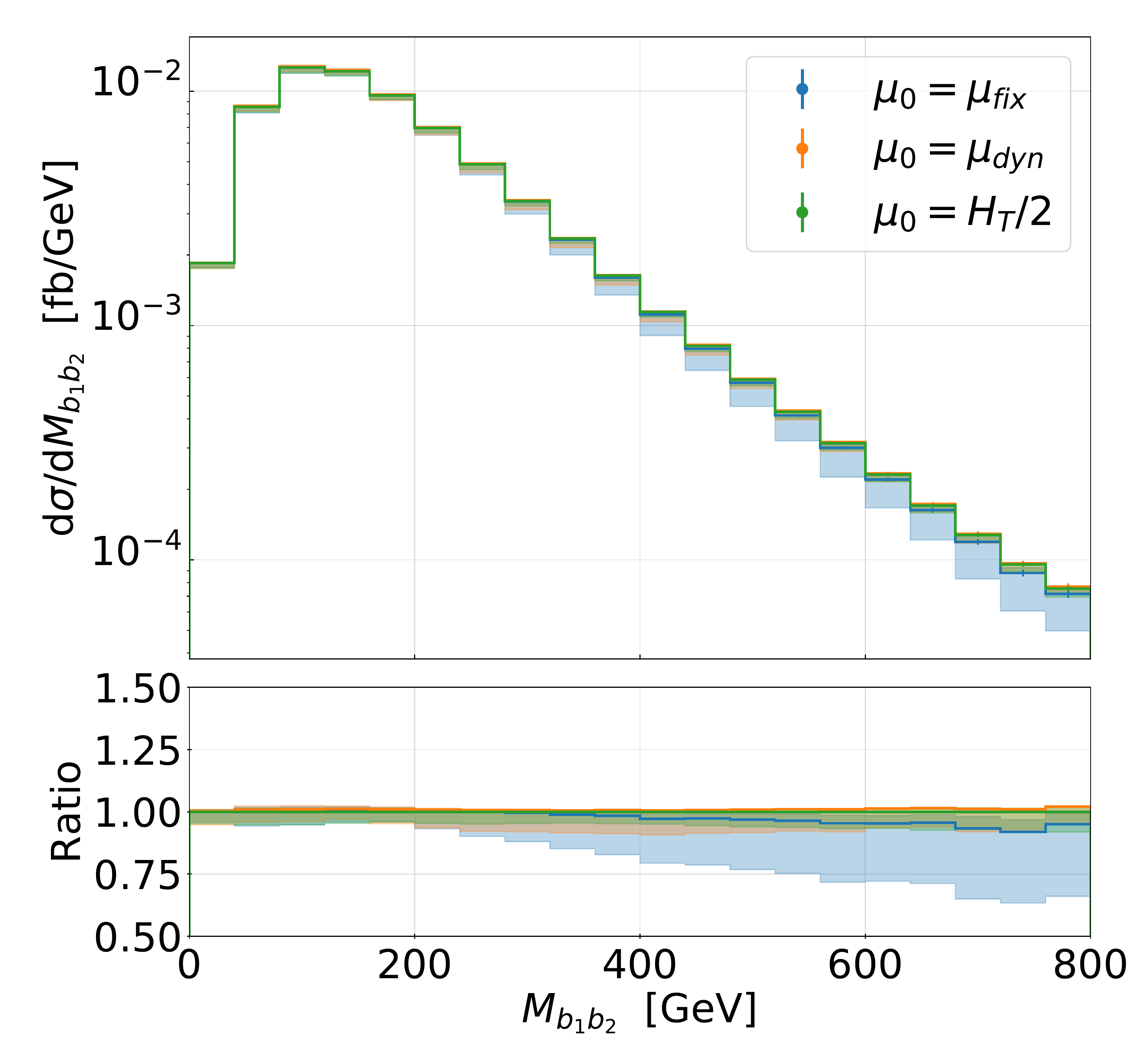}
\end{center}
\caption{\it \label{fig2} Differential distributions at NLO QCD for the observables
$M_{e^+\mu^-}$ and $M_{b_1b_2}$ for the $pp \to e^+ \nu_e \mu^-
\bar{\nu}_\mu b\bar{b} H +X$ process at the LHC with $\sqrt{s} = 13$
TeV. Results are shown for three different scale settings:
$\mu_0=\mu_{fix}$, $\mu_{dyn}$ and $\mu_0=H_T/2$.  The upper panels
show absolute predictions together with  the  corresponding uncertainty
bands resulting from scale variations. Also given are Monte Carlo
integration errors. The lower panels display the ratio to the $\mu_0 =
H_T /2$ case.}
\end{figure}

To emphasise  the need to use dynamic scale settings even more, we
present in Figure \ref{fig2} NLO differential cross-section
distributions for the invariant mass of two leptons $(M_{e^+\mu^-})$
and two $b$-jets $(M_{b_1b_2})$ for three different scale settings. In
addition to $\mu_0 = \mu_{fix}$ and $\mu_0=H_T/2$ we also plot
theoretical predictions for $\mu_0=\mu_{dyn}$ where
$\mu_{dyn}=\left(m_{T,\,t}\,m_{T,\,\bar{t}}\,m_{T,\,H}\right)^{1/3}$
and $m_{T,\,i}=\sqrt{m_i^2+p_{T,\,i}^2}$\,. We construct top and
anti-top momenta directly from their decay products. Scale
uncertainties are again displayed as uncertainty bands and the lower panels
show the ratio to the results obtained for $\mu_0 = H_T /2$. The two
dynamical scales are rather alike for $\mu_0$. The fixed scale
setting, however, is quite different already for the central value of
the scale. Furthermore, it results in significantly larger scale
uncertainties in the tails of  dimensionful distributions.
%
\begin{figure}[t!]
  \begin{center}
    \includegraphics[width=0.49\textwidth]{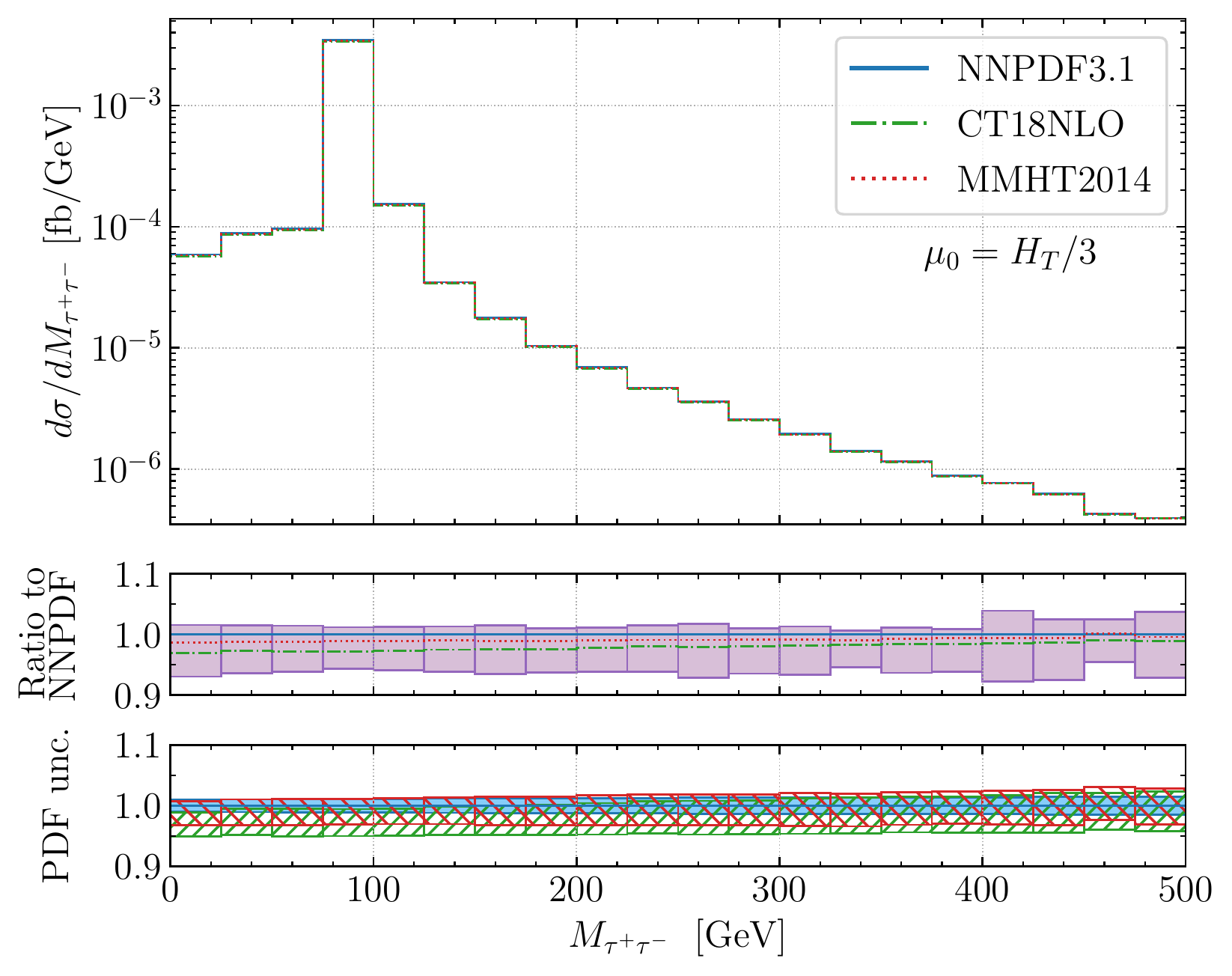}
  \includegraphics[width=0.49\textwidth]{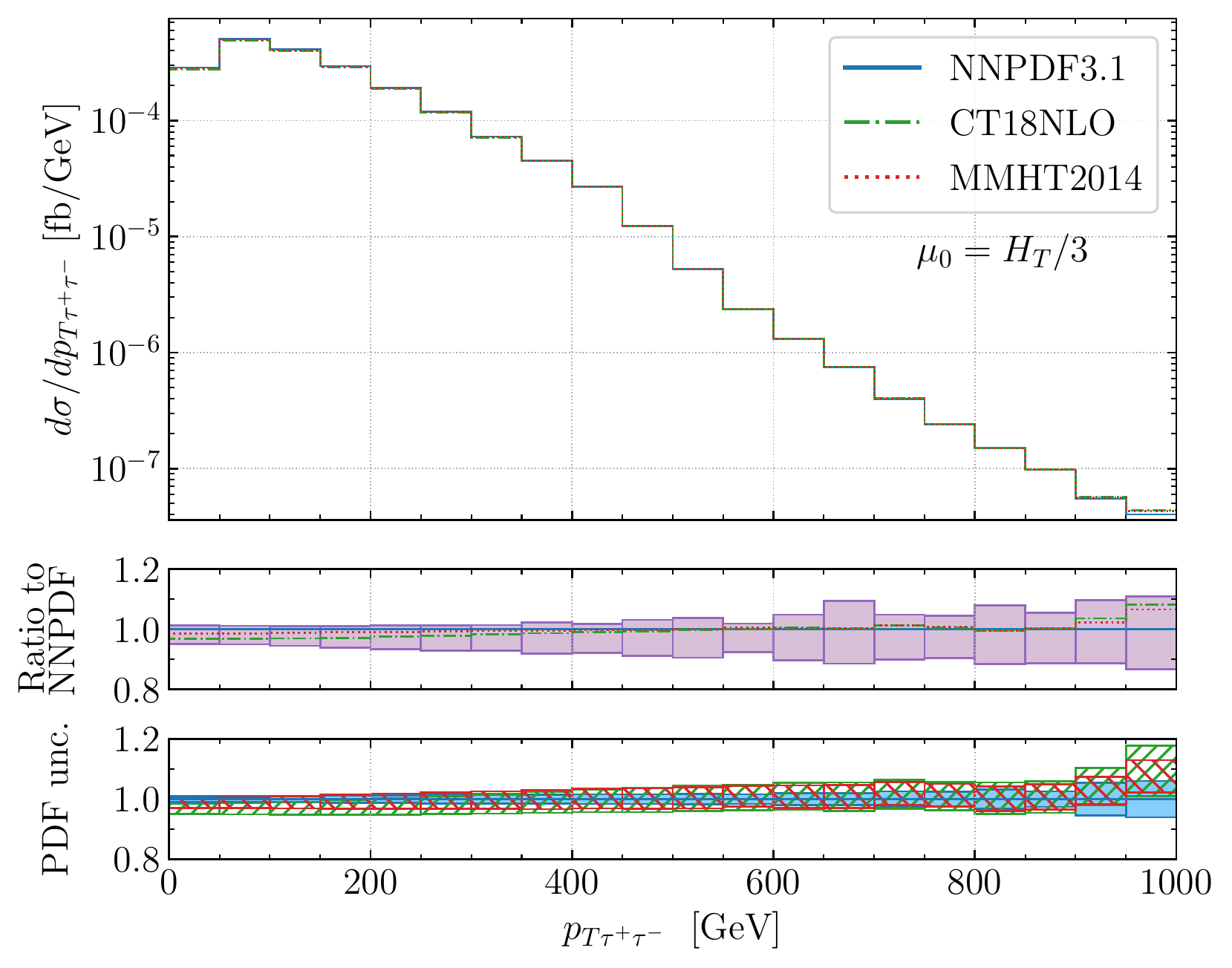}
  \end{center}
\caption{\it \label{fig3} Differential cross-section distributions for
$pp \to e^+ \nu_e \mu^- \bar{\nu}_\mu b\bar{b} \tau^+ \tau^- + X$ at
the LHC with $\sqrt{s}=13$ TeV as a function of $M_{\tau^+\tau^-}$ and
$p_{T \tau^+\tau^-}$. The upper panels show the absolute NLO QCD
predictions for three different PDF sets with
$\mu_R=\mu_F=\mu_0=H_T/3$. The middle panels display the ratio to the
result with the NNPDF3.1 PDF set as well as its scale dependence. The
lower  panels present the internal PDF uncertainties for each PDF
set. }
\end{figure}

Having examined the size of scale uncertainties for the differential
cross-section distributions, we turn our attention to the PDF
uncertainties. For this case, we concentrate on the $pp \to e^+ \nu_e
\mu^- \bar{\nu}_\mu b\bar{b} \tau^+ \tau^- + X$ process. All input
parameters and the list of selection criteria  are summarised 
in  Ref. \cite{Bevilacqua:2022nrm}. In Figure \ref{fig3} we
display the following two observables, the invariant mass of the
$\tau^+\tau^-$ system $(M_{\tau^+\tau^-})$ and the transverse momentum
of the $\tau^+\tau^-$ system $(p_{T,\, \tau^+\tau^-})$. We plot them
for the CT18 and MMHT14 PDF sets as well as for NNPDF3.1. Each plot
comprises three panels. The upper panel displays the absolute NLO
prediction for three different PDF sets at the central scale value,
$\mu_R=\mu_F=\mu_0=H_T/3$ where $H_T$ is defined this time as $H_T=
p_{T,\,b_1} + p_{T,\,b_2} + p_{T,\,e^+} +p_{T,\,\mu^-}+p_{T,\,\tau^+}
+p_{T,\,\tau^-} + p_T^{miss}$. The middle panel shows the NLO QCD
scale dependence band normalised to the NLO prediction obtained with
$\mu_0$ and the NNPDF3.1 PDF set. Also given is the ratio of NLO QCD
predictions generated for CT18 and MMHT14 to the result calculated with
the help of the NNPDF3.1 PDF set. The
lower panel  displays the internal PDF uncertainties for each PDF set
separately, which are also normalised to central NLO predictions with
NNPDF3.1. We observe that at the differential level,  the NNPDF3.1 PDF
uncertainties are very small and well below the corresponding
theoretical uncertainties due to scale dependence. When analysing the
internal PDF uncertainties for CT18 and MMHT14, we notice that they
behave similarly to each other and that their PDF uncertainties are almost a factor of 2
larger than those for NNPDF3.1. Nevertheless, for the selected 
PDF sets the internal PDF uncertainties are still within the
theoretical uncertainties due to scale dependence. Thus, the latter
are the dominant source of the theoretical systematics for the
differential cross-section distributions at NLO in QCD.
%
\begin{figure}[t!]
  \begin{center}
    \includegraphics[width=0.49\textwidth]{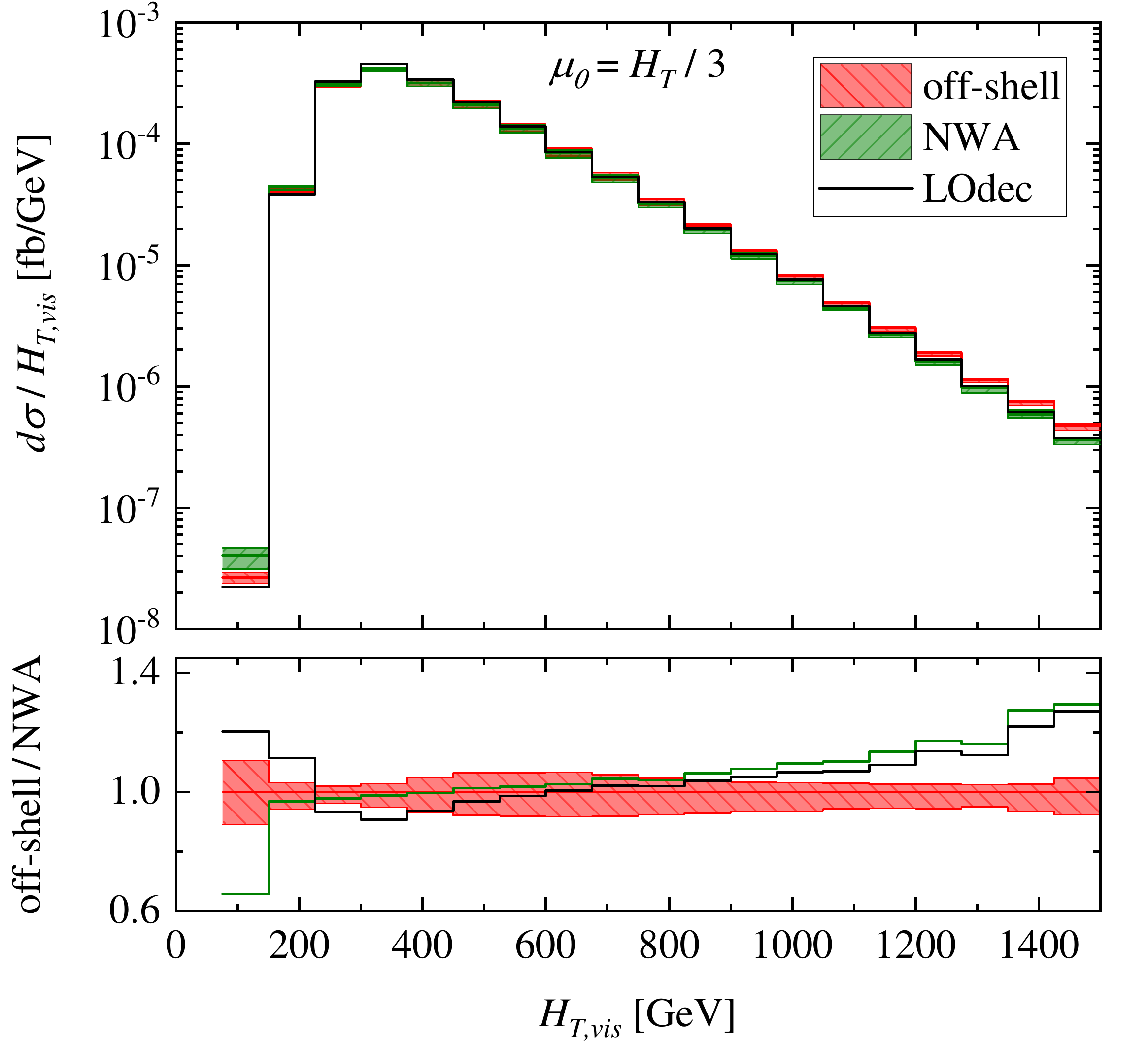}
    \includegraphics[width=0.49\textwidth]{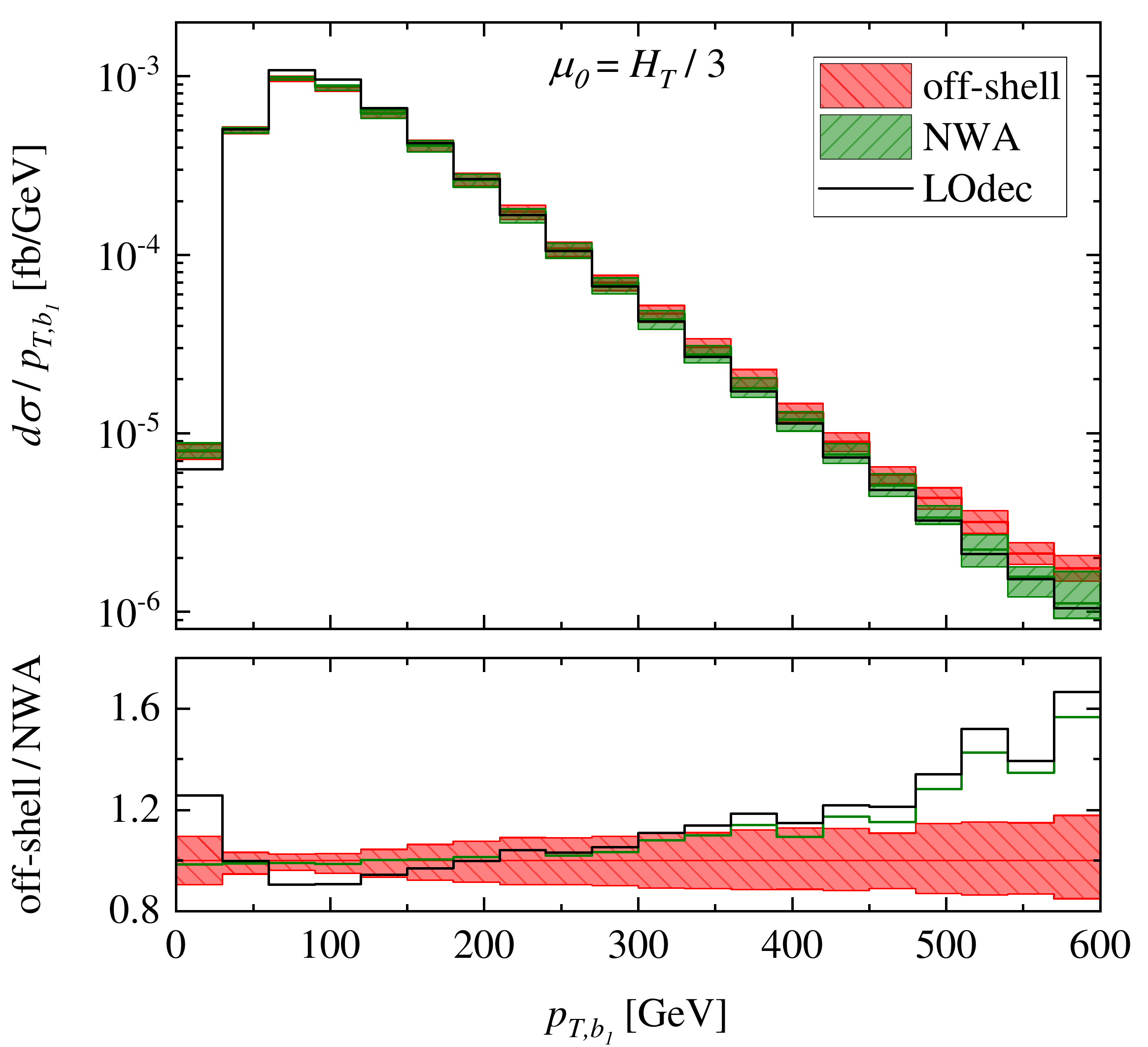}
\end{center}
\caption{\it \label{fig4} Differential cross-section distribution as a
  function of  $H_T^{vis}$ and $p_{T,\, b_1}$ for the $pp \to e^+
  \nu_e \mu^- \bar{\nu}_\mu e^+ \nu_e b\bar{b}+X$ process at the LHC
  with $\sqrt{s}=13$ TeV. NLO QCD
results for various approaches for the modelling of top quark
production and decays are shown. Additionally, theoretical
uncertainties as obtained from the scale dependence for the full
off-shell case are provided. Also plotted are the ratios of the full
off-shell result to the two NWA results.}  
\end{figure}

Finally, we examine the size of the non-factorisable corrections for
the $pp \to t\bar{t} B+X$ process.  To inspect them closely,  we compare
the NLO QCD results with the complete 
off-shell effects  to the calculations in the NWA. The NWA
results are divided into the following two categories: the full NWA
and the ${\rm NWA}_{\rm LOdecay}$ case. The full NWA comprises NLO QCD
corrections to $t\bar{t}W^\pm$ production and the subsequent
top quark decays, preserving at the same time top quark spin
correlations. The ${\rm NWA}_{\rm LOdecay}$ case contains the results
with NLO QCD corrections to the production stage only, whereas top
quark decays are calculated at LO.  For consistency, the NWA results
with top quark decays at  LO are calculated with 
$\Gamma_{t,\, {\rm    NWA}}^{\rm LO}$.

In general, for very inclusive observables the non-factorisable
corrections vanish in the limit $\Gamma/m \to 0$, which 
satisfies the following hierarchy
\begin{gather*}
  \frac{ \Gamma_Z}{m_Z} \,\, > \frac{ \Gamma_W}{m_W}
\,\, > \,\, \, \frac{\Gamma_t}{m_t}\,\,\, \gg\,\,\,
\frac{\Gamma_H}{m_H} \,,\,\,\,\,\, \\[0.2cm] 2.7\% \, >\, 2.6\% \, >\,
0.8\% \, \gg \, 0.003\% \,.
\end{gather*}
Thus, we can expect that the  complete off-shell effects are rather small
for integrated cross sections for $pp \to t\bar{t}B+X$
production. Indeed, such contributions turn out to be small in 
inclusive cross sections. However, they are strongly enhanced in some 
exclusive observables that play an important role in Higgs boson 
measurements  and new physics searches.

In the following, we present our findings for the $pp\to e^+ \nu_e \,
\mu^- \bar{\nu}_\mu \, e^+ \nu_e \, b\bar{b} +X$ process. All input
parameters, PDF sets and the cut selection that we have
employed can be found in Ref \cite{Bevilacqua:2020pzy}. In Figure
\ref{fig4}, we display differential cross-section distribution as a
function of $H_T^{vis}$ and $p_{T,\, b_1}$. The former observable is
defined as $H_T^{vis} = p_T(\mu^-) +p_T(e^+) + p_T(e^+) + p_T(j_{b_1})
+ p_T(j_{b_2})$. We plot theoretical results at NLO in QCD for the
following  three cases, the full NWA, NWA${}_{\rm LOdecay}$ and full
off-shell predictions. For the full off-shell case, we additionally display
the theoretical uncertainties as obtained from the scale dependence
since we are interested in effects that are outside the NLO
uncertainties bands. The upper plots show the absolute predictions at
NLO in QCD, whereas the bottom plots  display the ratios of the full
off-shell result to the two NWA results. We can observe that in the tails
of $H_T^{vis}$ and $p_{T,\, b_1}$,  full off-shell effects increase
and are well above the theoretical  scale uncertainties. Furthermore, at the beginning of both spectra we can
notice large discrepancies between the full NWA description and the
NWA${}_{\rm LOdecay}$ case. They are visible in the phase-space
regions that are currently scrutinised by ATLAS and CMS.  This
highlights the importance of the proper modelling of top quark decays
for the $pp \to t\bar{t}B+X$ process.

Overall, in the case of various (dimensionful) differential cross-section distributions for  the $pp \to t\bar{t}B+X$ process,
non-negligible full off-shell effects are present in various
phase-space regions. Substantial differences between the two versions
of the NWA results are additionally observed. Taking into account that, a priori, it is not possible to estimate the size of these
non-factorisable corrections and
which phase-space regions are particularly affected, a very careful
examination based on the full theoretical description should be
performed on a case-by-case basis. Let us just mention at this point
that  full off-shell
effects  also have a significant impact on the top quark mass extraction
\cite{Bevilacqua:2017ipv}, the charge asymmetry of the top quark and
its decay products \cite{Bevilacqua:2020srb} as well as the
calculation of signal strength exclusion limits in $t\bar{t}$
associated Dark Matter production \cite{Hermann:2021xvs}.

\section{Conclusions}

We have briefly summarised NLO QCD corrections to $pp \to
t\bar{t}B+X$, where $B=H,Z,W^\pm$. In  these computations, off-shell top
quarks and $W^\pm/Z$ have been described by  Breit-Wigner
 distributions.   Furthermore, double-, single- as well as non-resonant
contributions along with all interference effects have been
consistently incorporated at the matrix element level. We have presented
our results for the LHC Run II centre of mass system energy of
$\sqrt{s} = 13$ TeV for various differential cross-section
distributions. We employed  several scale settings and three different
PDF sets. Large shape distortions have been observed in the presence
of higher-order QCD effects. The non-flat differential ${\cal
K}$-factors  underline the importance of NLO QCD corrections for a
proper modelling of the $pp \to t\bar{t}B+X$ process
kinematics. Furthermore, we have observed that the fixed scale setting led
to perturbative instabilities in the TeV regions. The introduction of
the dynamical scale  stabilises the high $p_T$ tails and generally leads to smaller NLO QCD corrections as well as theoretical
uncertainties. In addition, the size of full off-shell effects has
been examined. At the differential level, large non-factorisable
corrections  up to $40\% - 60\%$ have been observed. Last but not
least, the differences between the full NWA and NWA${}_{\rm LOdecay}$
case  are substantial, especially in the low $p_T$ regions. The latter
phase-space regions are currently scrutinised by ATLAS and CMS. To
recapitulate, the non-factorisable NLO QCD corrections as well as
higher order QCD effects in top quark decays  significantly impact
$pp \to t\bar{t}B+X$ cross sections in various phase-space regions. For
these reasons, they should be included in future comparisons
between theoretical predictions and experimental data.

\acknowledgments
The author would like to thank the organizers of the $16^{th}$
Workshop on Elementary Particle Physics - {\it 
Loops and Legs in Quantum Field Theory} -  for a kind
invitation and very pleasant atmosphere during the conference.

The work was supported by the Deutsche Forschungsgemeinschaft (DFG)
under the following grant: 396021762 - TRR 257: {\it P3H - Particle
  Physics Phenomenology after the Higgs Discovery}. Support by a grant
of the Bundesministerium f\"ur Bildung und Forschung (BMBF) is
additionally acknowledged.

\end{document}